# Extracting Core Claims from Scientific Articles

Tom Jansen[1] and Tobias Kuhn[2]

[1,2] Vrije Universiteit Amsterdam, Faculty of Science, De Boelelaan 1105,
1081 HV Amsterdam, the Netherlands

[1] `t.d.jansen@student.vu.nl`
[2] `t.kuhn@vu.nl`

**Abstract.** The number of scientific articles has grown rapidly over the years and there are no signs that this growth will slow down in the near future. Because of this, it becomes increasingly difficult to keep up with the latest developments in a scientific field. To address this problem, we present here an approach to help researchers learn about the latest developments and findings by extracting in a normalized form core claims from scientific articles. This normalized representation is a controlled natural language of English sentences called AIDA, which has been proposed in previous work as a method to formally structure and organize scientific findings and discourse. We show how such AIDA sentences can be automatically extracted by detecting the core claim of an article, checking for AIDA compliance, and – if necessary – transforming it into a compliant sentence. While our algorithm is still far from perfect, our results indicate that the different steps are feasible and they support the claim that AIDA sentences might be a promising approach to improve scientific communication in the future.

## 1    Introduction

The number of scientific articles is rapidly growing [8]. With this overwhelming amount of information, the need for proper information extraction tools that only extract the relevant information is increasing too. There is so much textual information out there that scientists can easily get lost looking for the right information. This overload makes it hard and time-consuming for researchers to keep up with the latest developments in their scientific field. When finding the latest developments proves too difficult, duplicate research may be conducted leading to unnecessary work and research that is already conducted elsewhere. The field of text mining [1, 17] has played a crucial role in extracting relevant information from (scientific) literature. Text mining is primarily used for the decomposition and analysis of texts or textual information. Gener-ally, it refers to the process of extracting interesting – or relevant – information or patterns from unstructured documents [17]. Often used techniques include part-of-speech tagging, named entity recognition and sentiment analysis. Techniques like these pave the way for more complex and interesting text analyses. The extraction of information and knowledge from texts is such a complex step. Over the years, enormous progress has been made with respect to text mining in the areas of information retrieval, evaluation methodologies and resource construction especially in the biological domain [21]. Despite this progress, the problem of information overload has not been solved yet, and it seems that we need more than just text mining. To overcome the issue at hand and improve the way we share scientific findings, we might have to change scientific communication altogether. A possible solution would be that text mining and other machine learning approaches are supported by the way we publish scientific findings in the future. The concept of AIDA sentences is such an approach and offers a way to structure scientific findings and discourse [7]. AIDA sentences are a controlled natural language [6] of single independent English sentences that can represent scientific claims in a normalized and re-usable fashion. AIDA sentences are proposed as a tool for researchers to easily access and communicate research hypotheses, claims and opinions. The vision is that scientists will summarize their findings in AIDA sentences and interlink them with other claims, but to take into account the large body of existing literature, we need to apply text mining methods to integrate past and future research in a symbiosis of manual and automated work. In this paper, an attempt is made to start this integration of past and future research. To do this, we present a 3-fold approach to extract core claims from scientific articles to bootstrap the AIDA approach. Step one is to find the core sentence of a scientific article. Second, a check is performed to see whether the sentence complies with the AIDA rules. If it does not, a third and final step is undertaken that attempts to rewrite the core sentence into an AIDA sentence.

The rest of this paper is organized as follows. Section II presents background information and related work that may be helpful in achieving the goal of this paper. Here, the concept of AIDA sentences is explained in detail as well. Section III provides a more elaborate explanation of the approach. In Section IV the experimental framework that is used to train and test the developed algorithm is discussed. Results of these tests are provided in Section V, after which they are discussed in Section VI. Finally, the document is concluded in Section VII.

## 2 Background

Below, we introduce the concept of AIDA sentences, which plays a key role in our approach. Therefore, this will be addressed first. Closely related to the goal of this paper, and widely researched, is the topic of document summarization. This will be addressed here as well. Before we delve into document summarization, however, we take a look at related work with respect to keyword and keyphrase extraction – which is often an important step in document summarization.

## 2.1 AIDA Concept

AIDA sentences have been proposed as a tool for researchers to access and communicate scientific claims [7]. An AIDA sentence represents a single scientific claim that provides the core finding of an article. When such claims are extracted from articles and easily accessible, researchers can easily keep up with the latest developments. There are four requirements that need to be fulfilled by a sentence to be AIDA – each denoted by a single letter in AIDA. This structure will play a key role in this paper and its approach to extract scientific claims from articles. Here, we show two example sentences that comply with all the requirements to be AIDA sentences:

- Malaria is transmitted by mosquitos.
- The degree of hepatic reticuloendothelial function impairment does not differ between cirrhotic patients with and without previous history of SBP.

These sentences not only comply with all the requirements, they also display a clear statement of the scientific finding of a given article – embracing the bigger picture of AIDA sentences. In other words; AIDA sentences comply with a set of requirements, which make sure the sentence provides a single scientific claim in a normalized fashion. Ideally, such a sentence is written by the authors themselves. However, since this is often not enforceable (in particular for already written articles), they can be automatically extracted to get it started. The AIDA requirements, quoted from [7], are:

- **Atomic**: a sentence describing one thought that cannot be further broken down in a practical way
- **Independent**: a sentence that can stand on its own, without external references like "this effect" or "we"
- **Declarative**: a complete sentence ending with a full stop that could in theory be either true or false
- **Absolute**: a sentence describing the core of a claim ignoring the (un)certainty about its truth and ignoring how it was discovered (no "probably" or "evaluation showed that"); typically in present tense

The absoluteness criterion might sound suspicious at first, as uncertainty and the context of discoveries are crucial aspects of scientific findings. In fact, the AIDA approach does not deny the importance of these aspects, but assumes that they can be formally linked to an AIDA sentences, using models such as the ORCA model [19]. Therefore, it is argued that these aspects do not need to be present in the sentences. Together, these requirements should allow AIDA sentences to be re-used (e.g. another paper reproducing a claim) and interlinked (subsumption hierarchies, equivalence/relatedness, etc.), and thereby enable the efficient organization of scientific discourse. Furthermore, formal provenance and metadata (including scientific method, degree of confidence, source article, authors, timestamp, used data, etc.) can be attached to AIDA sentences with the nanopublication technique [11].

## 2.2 Keyword and keyphrase extraction

Keywords are often defined as the most important words of a scientific article. They comprise the subject(s) of an article and provide a concise summarization of a document. Because of this, they are an important feature for techniques such as document retrieval and topic search [20]. Whereas keywords are single word terms, keyphrases consist of multiple words (e.g. *keyword extraction*). Authors often include manually assigned keywords/keyphrases in their articles, but there are still many articles that lack any of the two. There are a number of approaches by which the extraction of keywords and keyphrases can be carried out. Broadly speaking, there are four methods: rule-based linguistic approaches, statistical approaches, machine learning approaches (supervised, unsupervised and semi-supervised) and domain specific approaches [16]. An – old-fashioned but still effective – statistical approach that is often used is TF-IDF. This feature is used to reflect the importance of a word to a document in a collection or corpus. TF-IDF stands for Term Frequency-Inverse Document Frequency and determines the importance based on the frequency of a word in a document versus the frequency of that word in the whole corpus. This simple algorithm efficiently categorizes relevant words [12]. Of the four methods, however, machine learning approaches are most prevalent in academic literature [3]. Early research into keyphrase extraction approaches this problem as a supervised learning task [18]. In that research, a genetic algorithm and a set of adjustable parameters were used for keyphrase extraction. A more recent study proposed an extended Term Frequency method to extract keywords [5]. Multiple features were used and were given weighting methods, after which an SVM model was used on the results for further optimization. In an unsupervised approach, another study extracts keyphrases based on the observation that keyphrases frequently contain multiple words but rarely standard punctuation or stop words [13]. Finally, research has shown that the highest frequency of keywords per noun in biomedical texts is found in the abstract [15]. Since the abstract is a short summary of the article, it makes sense this part contains a great deal of relevant information in a relatively small piece of text. However, it is also stated that other sections of biomedical texts are worthwhile to go through since they potentially host many keywords as well.

## 2.3 Document summarization

Like keyphrase extraction, automatic document summarization has received a lot of attention over the years. On the one hand, both are similar in the fact that they aim to determine the essence of a text or document. On the other hand, document summarization is more complex because it not only deals with words or phrases but whole sentences and larger bodies of text. There are two main methods of automatic summarization: extractive and abstractive. An extractive summary contains a set of sentences from the document, whereas an abstractive summary can contain material that is not present in the document but is constructed [2, 14]. A popular approach to summarize documents nowadays is based on graph representations. TextRank is a well-known graph-based ranking model that is used for both keyword and sentence extraction [10]. The im-portance of words and sentences is based on the relation between them within the constructed graph. In the graph, sentences are represented as vertices. The higher the number of relations of a vertex within the graph, the higher the importance of that vertex. Another recent research provides a comparison between an extractive and abstractive approach to document summarization [9]. The extractive approach consists of five steps of which the middle step is topic identification. A summary is generated containing sentences from the document that are considered most relevant. For the abstractive approach, the extractive summary is used as a basis and a word graph is generated that is integrated with that summary to create an abstractive summary. Results show that both approaches perform similar in terms of the information the summaries contain, but that the abstractive summary is more appropriate from a human perspective. Finally, a study performed a quantitative and qualitative assessment of 15 algorithms for sentence scoring [4]. Out of the 15 assessed algorithms, five methods showed the best performance: word frequency, TF-IDF, lexical similarity, sentence length and the TextRank score.

## 3 Approach

To achieve the goal of extracting a single claim from a scientific article, we employ a rule-based approach. While machine learning approaches, as introduced in the Background section, have shown to perform well for a variety of problems, they can be expected to require training data of considerable size to lead to satisfactory results for our problem domain. Such datasets are not yet present with respect to scientific articles and their core sentences. A rule-based approach is therefore a natural first approach that could facilitate the creation of training data to employ machine learning methods on this problem in the future. Our rule-based approach consists of three steps: (1) the core sentence of an article is extracted; (2) that sentence is checked for AIDA compliance; (3) when the sentence does not comply with the requirements for an AIDA sentence, an attempt is made to rewrite the sentence to satisfy the requirements. At the highest level, the process of retrieving a scientific claim from an article is shown in Figure 1.

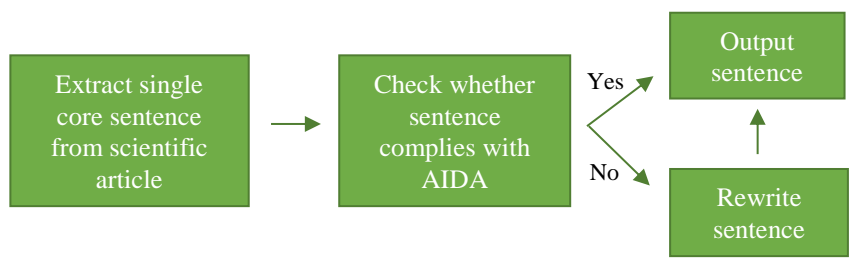

**Fig. 1.** Process of retrieving a scientific claim.

## 3.1 Extracting core sentences

First, the core sentence needs to be extracted from a scientific article. Core sentences are sentences that best describe the core idea or claim of an article. For the scope of this research, we have limited the search for the core sentence to the abstract of scientific articles alone. As denoted earlier, research has shown that the abstract contains the highest frequency of keywords for biomedical texts. To set up this research, we have therefore chosen to focus on the abstract. Furthermore, the assumption is made that a scientific article – or abstract in our case – contains one single sentence that best describes the core claim of that article. To find the core sentence, every sentence is rated with a score based on its content. When every sentence has received a score, the sentence with the highest score is extracted as the core sentence. The score of a sentence is based on four factors that will be briefly explained:

1. Whether the sentence matches a defined pattern
2. The number of 'core' and 'non-core words' in the sentence
3. Term frequency
4. Sentence length

Core sentences of scientific articles often start with a summarizing phrase like "*overall, this study reveals that*", "*in conclusion, these findings confirm*", "*the data suggest that*" etc. If an article contains a sentence with a similar structure or pattern, the chance is relatively high that this is the core sentence. To find sentences like these, we define these patterns by a regular expression. If a sentence is found that matches the regular expression, this sentence gets a relatively high score. The search continues, since several sentences can be structured like this.

Although not every sentence is structured in the (desired) way that is shown in the regular expression, they usually still contain words of the same scope. Verbs that are also used in the former regular expression like "*reveal*", "*confirm*", "*suggest*" potentially show that a scientific claim is coming. This also holds for nouns like "*study*", "*experiment*" and "*research*" and adjectives like "*overall*", "*altogether*" and "*collectively*". We stored all these and other similar words into a list. If a word from this list is present in a sentence, the score of the sentence is increased. Another list of words is also created, containing words that should not be present in a core sentence. These are words like "*sample*", "*survey*" and "*interview*". When one of these words is present, the score of the sentence is decreased.

Third, our algorithm creates a list that contains the ten most frequent terms of the article. As denoted in the Background section, term frequency and TF-IDF are often used and have proven to be successful. For this study, we have chosen to use term frequency alone. A list of the ten most frequent terms is compiled – excluding English stop words. For this particular step, the algorithm does not limit itself to the abstract but looks at the entire article to search for the most frequent words. Our algorithm assumes that words from this list are most important for the current article and therefore have a high chance of forming the claim of that article. When a frequent word is found, the score of the sentence in which the word is present is increased.

Finally, the length of a sentence is used as an indicator as well. Ideally, a core claim is not too long consisting only of a simple claim – like the example sentences shown above. In abstracts and articles, however, core sentences can be relatively long. To avoid very long sentences to be extracted, for example because they do contain a lot of important words, we penalize sentences that are longer than any of the labeled core sentences plus ten percent.

### 3.2 Checking AIDA rules

The second step of the algorithm is to check whether the extracted core sentence complies with the AIDA rules. For every individual rule, several checks are conducted. For three of the four rules (Atomic, Independent and Absolute) the most important check consists of a list of words that provide a strong indication for a violation of that particular rule.

An atomic sentence describes a single thought. We compiled a list of words that indicate, for example, a contradistinction or enumeration – implying more than one thought. In addition to this list, the algorithm checks whether the sentence consists of one or multiple clauses. Every sentence is parsed using a statistical parser. From the parsed tree, the algorithm derives whether the sentence consists of a single clause or several. If multiple clauses are present, the sentence is considered not atomic because it can potentially be broken down further.

A sentence is independent when there are no external references and the sentence can stand on its own. Our algorithm uses a list of phrases such as *"this study"*, *"this experiment"* and *"we"* to check for external references. When such a word is present, it means the author refers to something else than just the current sentence and therefore the sentence is not independent.

A sentence needs to be grammatically complete and correct and in theory verifiable to be declarative. For this requirement, the algorithm checks whether the sentence is grammatically well-formed. It checks whether it starts with a capital and ends with a full stop. To be grammatically correct a sentence needs to contain at least one verb and one noun. If any of these rules is violated, our algorithm flags the sentence as not declarative.

An absolute sentence describes the core of a claim without showing how that claim was discovered and whether its truth is (un)certain. The list of words for this rule contains words like *"probably"*, *"likely"*, *"suggest"* since they show (un)certainty. Furthermore, our algorithm looks for modal verbs in the sentence. Modal verbs are used to indicate modality – that is: certainty, probability, possibility, permission and so on. When a modal verb is present or any word in the compiled list, the sentence is considered not absolute. Finally, claims that are absolute are written in the present tense. The algorithm therefore looks for verbs that are written down in the past tense. When a verb in the past tense is found, the algorithm identifies the sentence as not absolute.

### 3.3 Rewriting non-AIDA sentences

When the core sentence is extracted and the outcome of the AIDA check is negative, the algorithm attempts to rewrite that sentence into one that does comply with the AIDA rules.

Our current algorithm does not cover the cases where an extracted sentence is not atomic or not independent. Rewriting these sentences are the most challenging cases, and our current study focuses on the low-hanging fruits. If a sentence is not declarative, however, the algorithm ensures that the rewritten sentence will start with a capital and ends with a full stop. When the extracted sentence is not absolute, a number of steps are taken. First of all, a regular expression is used to find and remove expressions like *"overall, the results show that"* and *"these findings indicate"*. Second, our algorithm looks for modal verbs and the verb that comes with this modal verb. These are then replaced with a single verb without modality. Verbs in past tense are dealt with in a similar fashion. In this case, the algorithm identifies whether the corresponding noun is in singular or plural form. When the form of the noun is retrieved, the verb is rewritten into the present tense. Finally, we remove words showing (un)certainty – using the same list we use to check for absoluteness.

After the requirements are processed, a final syntactic check is done to deal with double spaces and to ensure that all sentences start with a capital letter and end with a full stop.

## 4 Experimental Framework

Here, we provide a description of the framework that we used to train and test our algorithm. First, we will discuss the used modules and describe our data set. Second, the experiments that we conducted to train and test the algorithm are explained. The three steps of our algorithm that were specified in the previous section, are evaluated and therefore also described individually.

### 4.1 Modules and data set

Python 2.7 in combination with the NLTK library[1] is used for Natural Language Processing. NLTK modules are used to separate sentences, tokenize the sentences and to assign a Part-Of-Speech tag to every token. Furthermore, we use a statistical parser to parse sentences: the pyStatParser[2].

For our experiments, we make use of a total set of 250 articles from the PubMed Central FTP service[3]. From this FTP service, we took three archives (comm_use.0-9A-B.txt.tar, comm_use.C-H.txt.tar and comm_use.I-N.txt.tar). Out of these three archives, 250 articles were randomly picked from a wide variety of journals. These 250 articles were divided into a training set and a test set. The training set consists of 125 articles from six different journals. For the test set we ensured that there are no more than three articles of the same journal. This, to ensure that our algorithm would not be trained or tested domain specific but work equally well for all scientific domains. For every article, we handpicked the core sentence of the abstract beforehand to create our gold standard. In addition to these 250 random articles, we also used a list of 250 already checked unique sentences, taken from the result table[4] of existing work on AIDA sentences [7]. Out of the 250 checked sentences, 65 sentences were labeled as not compliant with all of the AIDA rules.

---

[1] http://www.nltk.org/
[2] https://github.com/emilmont/pyStatParser - accessed on 09/03/2017.
[3] ftp://ftp.ncbi.nlm.nih.gov/pub/pmc/oa_bulk/ - accessed on 10/03/2017.

The full code and other required files, including the gold standards and comprehensive result files, can be found on GitHub[5].

### 4.2 Experimental setup

Here, we describe for each individual step of our approach, how we trained and tested the algorithm.

The first step that is trained and tested is that of extracting core sentences. As mentioned earlier, for this research we focus on extracting the sentence from the abstract alone. Our algorithm uses a number of parameters to give sentences a score. These parameters were optimized using the training set of 125 random articles. All the parameters either increase or decrease the score of a sentence with a maximum of 50 points. During optimization, many different combinations were tried before ending up with the optimal configuration. After this, testing was done with our test set of 125 different random articles. We then compared the 125 extracted sentences from the test set with the created gold standard to assess performance.

Second, we check whether sentences comply with the AIDA requirements. We trained our algorithm using the 250 sentences from the table from existing work on AIDA sentences. To measure the performance of the algorithm we used the 125 sentences that our algorithm extracted from the used articles. Before they were checked by the algorithm, we created another gold standard by checking the sentences ourselves. After we labeled all the sentences, we ran the algorithm and compared the results for evaluation.

Finally, the last step of our algorithm is to rewrite the sentences that do not comply with the AIDA requirements. For this step, we used the 250 entries from existing work on AIDA sentences for training again. During training, the sentences that were not compliant with all the requirements were rewritten and manually evaluated to improve the algorithm. After training, we tested this part using the correctly extracted sentences from the 125 abstracts that are used in the previous two steps. Given that the algorithm is trained to rewrite core sentences into single claims, we have chosen to omit the extracted sentences that did not match our gold standard and are thus not a core sentence. To assess the performance of this step, we manually evaluated the rewritten sentences.

---

[4] https://github.com/tkuhn/nanopubstudies-supplementary/blob/master/botstudy/extract_eval-results.ods - accessed on 15/03/2017.
[5] https://github.com/TomJansen25/Extracting-Core-Claims/

## 5 Results

Below, we show the performance of the individual steps of our algorithm.

### 5.1 Extracting core sentences

Before running the algorithm on the abstracts of 125 random articles, all the core sentences were handpicked. Due to the use of plain text files from the PubMed Central archives, in some cases the algorithm extracts some formatting (mostly a header) along with the sentence. To accommodate this, we compare the extracted sentence with the labeled core sentence from our gold standard and check how similar they are. When they show a similarity of at least 85%, the extracted core sentence is considered similar enough to the labeled core sentence. This is then a correctly extracted core sentence. When extracting the core sentence from the abstract, 77 out of 125 sentences match the labeled core sentence from the gold standard. Furthermore, in the remaining 48 sentences, ten extracted sentences also showed a claim of the article but not the core claim. Overall, 87 sentences containing claims were extracted from 125 abstracts of which 77 were also labeled beforehand as the core sentence. Thus, in 69.6% of the cases, a sentence is extracted from the abstract that contains a claim and in 61.6% of the cases, the perfect core sentence is extracted from the abstract.

### 5.2 Checking sentences for AIDA compliance

The set of 125 extracted sentences from the previous step is used to test the checks for AIDA compliance. The performance of the algorithm is measured using a set of four metrics: precision, recall, F-measure and accuracy. The diagram below shows an overview of the four metrics for every individual check and the AIDA check as a whole. Salient numbers will be discussed in the next section.

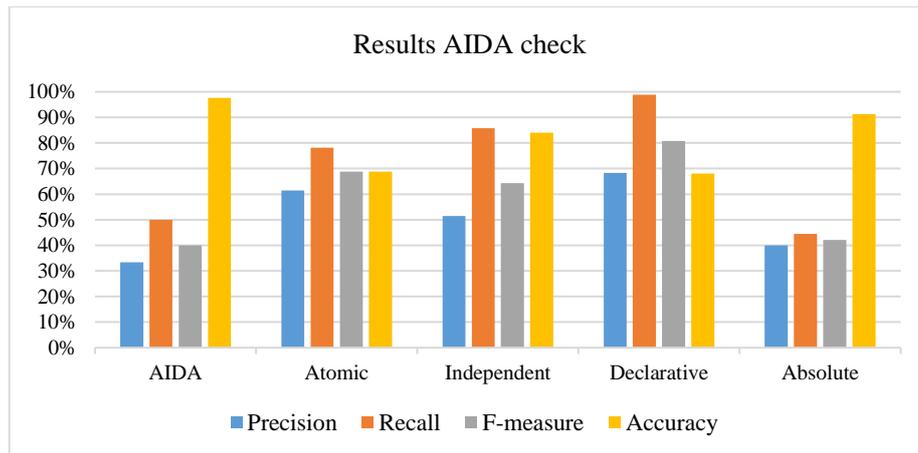

**Fig. 2.** Results of the AIDA sentence check.

## 5.3 Rewriting AIDA sentences

Out of the 87 extracted sentences that contained a claim, not a single one complied with all the requirements of being an AIDA sentence. Table 1 shows the compliancy of the 87 extracted sentences and the rewritten sentences according to our created gold standard.

| Requirement | Percentage of compliant sentences before rewriting | Percentage of compliant sentences after rewriting |
|---|---|---|
| **AIDA** | 0.00% | 28.74% |
| **Atomic** | 44.83% | 48.28% |
| **Independent** | 12.64% | 57.47% |
| **Declarative** | 74.71% | 90.80% |
| **Absolute** | 2.23% | 56.32% |

**Table 1.** AIDA compliance before and after rewriting.

As comes forward from Table 1, the resulting core sentences are improved in many ways after the algorithm has rewritten them. Again, these sentences were manually checked for AIDA compliancy and not by the algorithm to ensure a correct evaluation. Rewriting the sentences shows the best results regarding the independence and absoluteness requirements. Prior to rewriting the sentences, only 2.23% of the sentences was absolute and 12.64% independent. After rewriting the sentences, 56.32% of the sentences comply with the absoluteness requirement and 57.47% with the independence requirement. Overall, 25 out of 87 sentences comply with all the rules after rewriting, meaning that 28.74% of the sentences is correctly rewritten into AIDA sentences. Figure 3 shows an example of a completely correctly extracted and rewritten sentence.

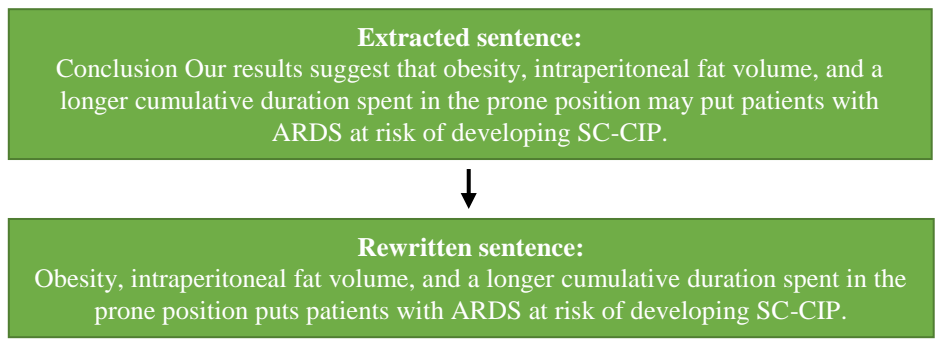

**Fig. 3.** An example of a completely correctly extracted and rewritten sentence.

# 6 Discussion

## 6.1 Extracting core sentences

The results of extracting a core sentence from the abstract are promising: in 69.6% of the cases, the algorithm picks a sentence from the abstract that contains a claim. Moreover, in 61.6% it picks the sentence that best describes the core claim. However, when taking into account that an abstract is usually 100-150 words long, it puts the results in perspective. Out of the approximately 10-15 sentences, picking the correct sentence should not be too difficult. For this research, we made the assumption that a scientific article or abstract consists of one core sentence. In the abstract alone, this also often is the case. Sometimes, however, the claim of an article is expressed in two or even more sentences. In some of our 125 random abstracts, multiple sentences showed a claim that was derived from the conducted research. Currently, our algorithm fails to recognize this and picks only the sentence with the highest score. For future improvements, the algorithm could be improved by including the possibility to extract multiple sentences or claims. Building upon this, more recent articles sometimes contain a highlight section that can aid in both the extraction of core sentences but also in the goal of communicating scientific discourse. This section consists of a number of core sentences that show the key findings of the article. Because this section is only present in a minority of the articles, we decided not to look at it but focus on the abstract. In the future, however, this section can be a great help in communicating scientific findings conform the concept of AIDA sentences.

## 6.2 Checking sentences for AIDA compliance

Looking at Table 2, the first thing that stands out is the big difference between the accuracy and the other three metrics in the overall AIDA check. This is, in part, due to the fact that out of all extracted sentences, only two comply all the AIDA requirements. Because the algorithm classifies most sentences as not compliant, the accuracy is high. However, when looking at the other three metrics, it shows that the algorithm has a lot of room for improvement. We will discuss the performance of every requirement individually to show performance and potential improvements.

First of all, results of the atomicity check show that all the metrics are relatively equal, with recall being a bit higher than the other three. For this check, the algorithm looks for certain words or phrases and whether the sentence consists of multiple clauses. If it consists of multiple clauses, the sentence can be broken down further and is therefore not atomic. Even though this is done, there are still many cases in which a sentence can be broken down but this is not recognized by the algorithm. Sentence parsing is done with a statistical parser that is not always correct. Due to this, multiple clauses are not always recognized. Improvements for this check can be made with respect to both parsing and the inclusion of more rules.

Second, for the independence check it is evident that precision is lower than the other metrics. With precision being less than the other metrics, it means that relatively many sentences are classified as independent when in fact they are not. To check for independence we only use a list of words and phrases that denote external references. Clearly, this list is not comprehensive enough and to improve this check, this list could be extended but other methods should also be investigated.

Striking from the metrics of the declarativeness check is that the recall is close to 100%, whereas the precision is just below 70%. The current checks for this requirement are minimal because it is hard to check whether a sentence is grammatically correct using a rule-based approach. Because of this, our algorithm classifies almost every sentence as declarative. Therefore, nearly every sentence that truly is declarative, is also classified as such. However, a great deal of sentences that are not declarative are still labeled as declarative – explaining why precision is relatively low. To improve this check, the 'understanding' of sentences should be improved which may prove hard for a rule-based approach like the one we use.

Finally, looking at the metrics of the absoluteness requirement, a similar thing is observed as with the AIDA check as a whole. Out of the 125 extracted sentences, only nine sentences were absolute. Because the algorithm (correctly) labels most of the sentences as not compliant, accuracy is high – over 90%. Precision and recall, however, are lower because the algorithm often fails to recognize that the sentence shows how something was discovered. In these cases, the sentences are classified as compliant when they are not. This check, like others, searches for pre-defined words and patterns that indicate a violation of the requirement. Extending this list could improve the performance of this check.

Overall, the metrics show that most of the checks can be improved in many ways. Including more rules could potentially improve the performance of the algorithm but only up until a certain point. However, our metrics also show the limitations of a rule-based approach. For future improvements, the inclusion and combination with other approaches should be investigated as well.

### 6.3 Rewriting AIDA sentences

The results reveal that the algorithm performs rather well when a sentence is not independent or not absolute. These two requirements often go hand in hand. As mentioned before, core sentences often start with an introducing phrase like "*this study suggests that*". This shows both how the results are retrieved (through this study) and is an external reference (to the study). Therefore, these two requirements are closely related. Sentences that are not absolute or not independent are often very well rewritten. As discussed in the Approach, especially when a sentence is not absolute, several things are checked and many changes can be brought about. Probably the most important check (and change) is that of the pattern matching with the regular expression. The introducing phrase mentioned earlier is often present in a core sentence and can therefore in many cases also be removed. For the algorithm to work even better, this pattern matching could be further refined.

The performance of the other two requirements, on the other hand, shows room for improvement. Our algorithm does not even try to deal with sentences that are not atomic and when a sentence is not declarative, some very simple measures are taken. First of all, the algorithm makes sure that the sentence will start with a capital and end with a full stop. Furthermore, headers that are extracted together with the sentence are removed as well to ensure declarativeness. With respect to the atomic requirement, the emphasis was placed on trying to remove clauses from a sentence during training. When sentences are not atomic, it may be because they consist of multiple sub clauses that are irrelevant. However, some taken measures showed to do more harm than good to a lot of sentences and therefore none of the measures were actually implemented. Therefore, when a sentence is not atomic, our current algorithm gives up and does not attempt any rewriting.

## 7  Conclusion

We presented an attempt to extract core claims from scientific articles. Our algorithm extracts the core sentence of the abstract, checks it for AIDA compliance, and tries to rewrite it in the negative case. In 69.6% of the cases, a sentence is extracted from the abstract that contains a claim from the article. Moreover, in 61.6% of the cases, the perfect core sentence is extracted from the abstract. Evaluating the checks for AIDA compliance was done both individually per requirement and for AIDA as a whole. Because only a few sentences complied with all the AIDA requirements, the accuracy of this check was very high (97.6%) but the F-measure rather low (40.0%). Finally, rewriting a sentence into a fully compliant AIDA sentence was done successfully 28.7% of the time. Overall, these results show that extracting AIDA sentences with a rule-based approach proves difficult but has great potential. With improvements and modifications in the algorithm, the extraction of core claims from scientific articles may become more accurate and efficient in the future. A critical mass of interlinked AIDA sentences made available with text mining might encourage authors to write AIDA sentences for their own findings and to interlink them with claims made by others. This, in turn, will allow us to automatically organize and aggregate scientific findings and discourse, and finally make it easier for researchers to keep up with the latest developments in their scientific field.